\documentclass[12pt,preprint]{aastex}
\usepackage{graphics}
\usepackage{subfigure}
\usepackage{amsmath}
\usepackage{multirow}
\usepackage{multirow}
\usepackage{color}

\newcommand{\smyr}{M$_\odot$ yr$^{-1}$}
\newcommand{\vkm}{km s$^{-1}$}
\newcommand{\degree}{$^{\circ}$}
\def\scaleboxs#1#2{}
\def\Jyb{Jy beam$^{-1}$}
\def\vsys{v_\textrm{\scriptsize sys}}
\def\Rin{R_\textrm{\scriptsize in}}
\def\Rout{R_\textrm{\scriptsize out}}
\def\arcs#1{$#1''$}
\def\arcsa#1#2{$#1^{\prime\prime}_{^\textrm{.}}#2$}
\def\leftblank#1{}
\def\ra#1#2#3#4{#1^\mathrm{h} #2^\mathrm{m} #3^\mathrm{s}_{^\textrm{.}} #4}
\def\dec#1#2#3#4{#1\degr #2\arcmin #3^{\prime\prime}_{^\textrm{.}}#4}

\begin{document}
\title{An Envelope Disrupted by a Quadrupolar Outflow
in the Pre-Planetary Nebula IRAS19475+3119}
\author{Ming-Chien Hsu\altaffilmark{1,2} and Chin-Fei Lee\altaffilmark{1}}
\altaffiltext{1}{Academia Sinica Institute of Astronomy and Astrophysics,
P.O. Box 23-141, Taipei 106, Taiwan; mchsu@asiaa.sinica.edu.tw, cflee@asiaa.sinica.edu.tw}
\altaffiltext{2}{Department of Physics, National Taiwan University, Taipei 10617,
Taiwan}

\begin{abstract}
IRAS 19475+3119 is a quadrupolar pre-planetary nebula (PPN), with two
bipolar lobes, one in the east-west (E-W) direction and one in the
southeast-northwest (SE-NW) direction. We have observed it in CO J=2-1 with
the Submillimeter Array at $\sim$ \arcs{1} resolution. The E-W bipolar lobe
is known to trace a bipolar outflow and it is detected at high velocity. The
SE-NW bipolar lobe appears at low velocity, and could trace a bipolar
outflow moving in the plane of the sky. Two compact clumps are seen at low
velocity around the common waist of the two bipolar lobes, spatially
coincident with the two emission peaks in the NIR, tracing dense
envelope material. They are found to trace the two limb-brightened edges of a slowly
expanding torus-like circumstellar envelope produced in the late AGB phase.
This torus-like envelope originally could be either a torus or a spherical
shell, and it appears as it is now because of the two pairs of cavities
along the two bipolar lobes.  Thus, the envelope appears to be disrupted by
the two bipolar outflows in the PPN phase.
\end{abstract}


\section{Introduction\label{Sec:Intro}}

\leftblank{A star in the AGB phase loses mass 
at a rate of $10^{-7} - 10^{-4}$ \smyr\ in the form of an isotropic
stellar wind, the so-called AGB wind. The ejected material forms a slowly 
expanding (10-20 \vkm) spherical circumstellar envelope (CSE) around 
the star, which is often seen as a roughly round halo in the optical 
\citep{Chu87,Sahai01}. After some 10,000 years such mass loss 
eventually depletes the hydrogen envelope of the AGB star and 
exposes its electron-degenerate carbon-oxygen core, leading to an increase in
the effective temperature of the star and the termination of the mass loss. 
The star then leaves the AGB phase and initiates a fast evolution to the left 
in the Hertzsprung-Russell diagram, i.e., increasing in temperature at an
approximately constant luminosity \citep{Blocker95}. When the star reaches 
temperature $\gtrapprox$ 30,000 K, the stellar UV photons with energy above 
the Lyman limit (13.6 eV) will be strong enough to photoionize the surrounding 
material (the former AGB CSE). This is the beginning of the PN phase, which 
is recognized by the emergence of strong recombination lines of hydrogen
(e.g., H$\alpha$) and helium collisionally excited lines of metals \citep{Kwok00}.
The phase through which a nebula around a post-AGB 
star evolves into a PN is referred to as the PPN phase.}

Pre-planetary nebulae (PPNe) are objects in a transient phase between the 
asymptotic giant branch (AGB) phase and the planetary nebula (PN) phase in
the stellar evolution of a sun-like star. 
The PPN phase is short with a lifetime of up to thousands of years; yet it closes 
the gap between the two morphologically very different phases
- the AGB phase and the PN phase. The former exhibits roughly spherical AGB
circumstellar envelopes (CSEs) expanding radially at 10--20 \vkm{},
comparable to the escape velocity of the AGB stars. Yet the latter exhibits
diverse morphologies, with a significant fraction having highly collimated
fast-moving bipolar or multipolar (including quadrupolar) outflow lobes
\citep{Balick02,Sahai04}.
In the past decades, high spatial resolution imagings of PPNe exhibit an
even richer array of shapes, including multiaxial symmetries
\citep{Balick02}, suggesting that the aspherical symmetry in the PN phase
must have started in the PPN phase \citep{Sahai04}. It seems that collimated
fast winds or jets are required to shape the PPNe and young PNe
\citep{ST98,Soker00,Lee03,Soker04,Sahai04}. In particular, the departure
from the spherical symmetry could result from an interaction of the jets
with the preexisting AGB CSEs. Also, multipolar and/or
point-symmetric morphologies could result from multiple ejections and/or
change in ejection directions
of the jets, respectively (Sahai et al. 2007, hereafter \citet{Sahai07}).

PPNe and PNe are often seen with a dense equatorial waist structure that is
often called a torus \citep[e.g., ][]{Kwok98, Su98, Volk07, Sahai08}. The
origin of the torus is unknown, and it may be due to either an equatorially
enhanced mass loss in the late AGB phase or an interaction between an
underlying jet and the spherical AGB wind \citep{Soker00}.
Tori are common in bipolar PPNe, but less common in multipolar PPNe.
It is thus important to study how a torus can be formed 
in multipolar PPNe and its relation with the jet.

The PPN IRAS 19475+3119 is a quadrupolar nebula with two collimated bipolar
lobes, one in the east-west (E-W) direction 
and one in the southeast-northwest (SE-NW)
direction \citep{Sahai07}.
In the optical image of the Hubble Space Telescope (HST),
the bipolar lobes appear limb-brightened, suggesting that
they are dense-walled structures with tenuous interiors or cavities
\citep{Sahai07}, and that they trace two bipolar outflows produced by two
bipolar post-AGB winds (S{\'a}nchez-Contreras et al. 2006, hereafter 
\citet{Sanchez06}; \citet{Sahai07}).
Its central star, HD 331319, is an F3 Ib (T$_{eff}\sim$
7200K) supergiant star classified as a post-AGB object based on the
elemental abundance analysis
\citep{Kloch02}. The distance is $\sim$ 4.9 kpc by assuming a total
luminosity of 8300 L$_\odot$, appropriate for a post-AGB star with a
mass of 0.63 M$_\odot$ \citep{Hrivnak05}. Using this luminosity and the
effective temperature, the central star is estimated to have a radius of 58
R$_\odot$.  This PPN has a detached circumstellar envelope based on its
double-peaked spectral energy distribution \citep{Hrivnak99}.  Near-infrared
(NIR) images of this PPN show a nebula extending in the
SE-NW direction and a puzzling two-armed spiral-like structure
\citep{Gled01}. The NIR nebula is the counterpart of the SE-NW bipolar lobe
in the optical. The spiral-like structure is point-symmetric and could
result from an interaction of the mass-losing star with a binary companion
\citep{Gled01}.

Single-dish CO spectra toward this PPN show two components, a strong line
core from the circumstellar envelope and weak wings from fast moving gas
\citep{Likkel91,Hrivnak05}. At $\sim$ \arcs{2} resolution, the two
components are seen in CO J=2-1 as a slowly expanding shell and a fast
bipolar outflow that is aligned with the E-W bipolar lobe in the optical
\citep{Sanchez06}. At $\sim$ \arcsa{0}{6} resolution, a ring-like envelope
is seen mainly around the E-W bipolar lobe (Castro-Carrizo et al. 2010,
hereafter \citet{Castro2010}). Here we present an observation
at $\sim$ \arcs{1} resolution in CO J=2-1 for this PPN.
Our observation shows an expanding torus-like envelope 
around the common waist of the two bipolar lobes and
that the structure and kinematics of the envelope
are consistent with the envelope being disrupted by the
quadrupolar nebula, or the two bipolar outflows.

\section{Observations}

Interferometric observation of IRAS 19475+3119 in CO J=2-1 was carried out using the Submillimeter Array (SMA)
\citep{Ho04} in the extended configuration on 2009 August 29. 
Seven antennas were used with baselines ranging from 44 to 226 m. The SMA
correlators were set up to have a channel spacing of 0.2 MHz ($\sim$ 0.26
\vkm) for CO J=2-1. The total duration of the observation (including the
integration time on the source and calibrators) was $\sim$ 11.5 hr with a
useful duration of $\sim$ 5.5 hr on the source.

The calibration of the data was performed using the MIR software package. 
Passband and flux calibrators were 3C84 and Uranus, respectively. 
The gain calibration was done with the quasar 2015+371, whose flux was 
estimated to be 3.05 Jy at 230 GHz, about 20\% higher than that reported from 
the light curve measurements of the SMA around the same period of time.
Each integration cycle includes 4 minutes on 2015+371 and 20
minutes on the source. 
The calibrated visibility data were imaged with the MIRIAD package.
The dirty maps that were produced from the calibrated visibility data
were CLEANed,
producing the CLEAN component maps.
The final maps were obtained by restoring the CLEAN component
maps with a synthesized (Gaussian) beam fitted to the main lobe of the dirty 
beam. With natural weighting, the synthesized beam has 
a size of \arcsa{1}{16}$\times$\arcsa{1}{01} 
at a position angle (P.A.) of $-$63\degree{}.
The rms noise levels are $\sim$ 0.1 \Jyb{} in the 
CO channel maps with a channel width of 1 \vkm{}.
The velocities of the channel maps are LSR.

\section{Results}

In the following,
our results are presented in comparison to
the HST image of \citet{Sahai07} and the NIR image of
\citet{Gled01}. Note that 
the HST image adopted in \citet{Sanchez06} was incorrectly 
rotated by $\sim$ 10\degree{} clockwise.
The systemic velocity in this PPN is found to be 18.25$\pm$1 \vkm{}
from our modeling in Section \ref{ssec:model} \cite[see also][]{Sanchez06}.
The integrated line profile of CO shows that most emission comes
from within $\pm15$ \vkm{} from the systemic velocity.
Compared with previous single-dish and interferometric observations in
the same emission line \citep{Sanchez06}, our interferometric observation at 
high resolution filters out most of the extended emission at low velocity,
allowing us to study compact structures, such as a torus and collimated
outflows, around the source.

\subsection{A torus-like circumstellar envelope}\label{sec:res_torus}

At low velocity (within $\pm$ 10 \vkm{} from the systemic velocity), the CO
map (Fig. \ref{12CO+HST}) shows two compact clumps on the opposite sides of
the source, one to the northeast (NE) at P.A. $\sim$ 15\degree{} and one to
the southwest (SW) at P.A. $\sim$ 195\degree{}, at a distance of $\sim$
\arcs{1} (4900 AU) from the source. These two clumps
can be seen in the 7 central velocity channels
from 8 to 29 \vkm{} (Fig. \ref{Fig:Cha12CO}) and have also been seen
at higher resolution in \citet{Castro2010} as well. They are seen around the
common waist of the two bipolar lobes, spatially
coincident with the two emission peaks in the NIR,
tracing the dense envelope
around the source. The center of the two clumps, however, is shifted by
\arcsa{0}{15} to the east of the reported source position, which is
$\alpha_{(2000)}=\ra{19}{49}{29}{56}$,
$\delta_{(2000)}=\dec{+31}{27}{16}{22}$ \citep{Hog1998}.
This position shift is only $\sim$ 15\% of the synthesized beam and thus can
be due to the position uncertainty in our map. Also, there could be uncertainty
in the reported source position. Therefore, in this paper, the
center of the two clumps is considered as the source position. 
Faint protrusions are also seen along the two bipolar lobes at low velocity,
extending to the east and west, as well as to the southeast and northwest.
These faint protrusions are most likely to be from the quadrupolar
lobes, and thus unlikely to be from the dense envelope itself.

The structure and kinematics of the two clumps can be studied with
position-velocity (PV) diagrams cut along four different P.A., from the one
aligned with their peaks (cut A with P.A.=15\degree{}) to the one
perpendicular (cut D with P.A. = $-75$\degree{} ), through the two in
between (cut B with P.A.=$-10$\degree{} and cut C with
P.A.=$-30$\degree{} ). The PV diagrams along these cuts shows a similar
ring-like PV structure with less and less emission around the systemic
velocity as we go from cut A to cut D (Figures \ref{12CO-PVobs}a-d),
indicating that the two clumps trace the two limb-brightened edges of a
torus-like envelope at a small inclination angle (i.e, close to edge-on),
either radially expanding or collapsing, with the equator aligned with cut
A. In addition, the torus-like envelope has a thickness with a half opening
angle (i.e., subtended angle or see next section for the definition) of
20\degree{}--45\degree{}, measured from its equator. The two high-velocity
ends of the ring-like PV structure are from the farside and nearside of the
torus-like envelope projected to the source position. The emission
there is bright, indicating that the material in the farside and nearside
must be projected to the source position, and this requires the torus-like
envelope to have an inclination angle smaller than its half opening
angle.
 Figure \ref{spectrum} shows the spectrum integrated over a
region with a diameter of $\sim$ \arcsa{2}{5} around the source.
It includes most of the torus-like envelope but excludes
the protrusions along the two bipolar lobes.
It shows a
double-peaked line profile with the redshifted peak at $\sim$ 28 \vkm{}
slightly brighter than the blueshifted peak at
$\sim$ 10 \vkm{}. Since the temperature of the torus-like envelope is
expected to drop away from the central star
\citep{Kwan82}, the
emission from the farside is expected to be brighter than that from the
nearside because of the absorption in the nearside. Therefore, the
redshifted peak in the line profile is from the farside and the blueshifted
peak is from the nearside, implying that the torus-like envelope is
expanding at $\sim$ 10
\vkm{}, in agreement with it being the circumstellar envelope
produced in the AGB phase. The PV diagram cut perpendicular to the equator,
i.e., cut along P.A.=-75\degree, shows that the redshifted emission shifts
to the west and the blueshifted emission shifts to the east at low velocity
(Fig. \ref{12CO-PVobs}d). For an expanding torus, this indicates that
the torus is inclined, with the nearside tilted to the east.

\subsection{Outflows}\label{sec:res_outflows}

At high velocity (beyond $\pm$13 \vkm{} from the systemic velocity), CO
emission is seen around the E-W bipolar lobe seen in the optical, emerging
from the two opposite poles (or holes) of the torus-like envelope (Fig.
\ref{12CO+HST}). This E-W bipolar lobe is already known to trace a
fast-moving bipolar outflow \citep{Sanchez06}. The PV diagram (Fig.
\ref{12CO-PVobs}e) along the outflow axis (P.A.=88\degree; cut E in Fig.
\ref{12CO+HST}a) shows that the outflow velocity increases from the base to
the tip at $\sim$
\arcs{4} ($\sim$ 20000 AU), as found in \citet{Sanchez06}.
This is why the CO emission at low-velocity is seen near the source
forming the faint protrusions along the E-W bipolar lobe, as mentioned in the
previous section. The radial velocities are $\sim$ 30 \vkm{} at
the tip of the red-shifted lobe and $\sim$ 24 \vkm{} at the tip of the
blue-shifted lobe, with a mean velocity of $\sim$ 27 \vkm{}.

The PV diagram also shows a hint of bifurcation in velocity along the
outflow axis, especially for the red-shifted lobe in the east [see the dark lines in
Figure \ref{12CO-PVobs}e and also Figure 20 in \citet{Castro2010}]. This suggests that the outflow lobe is a hollow,
shell-like cavity wall, as suggested in \citet{Sahai07}. For example,
in the case of the redshifted lobe, 
the emission with less red-shifted velocity is from the
front wall, while the emission with the more red-shifted velocity is from
the back wall. 
The outflow has the shell-like
structure, likely resulting from an interaction of
a post-AGB wind with the pre-existing circumstellar
medium, as proposed in \citet{Sanchez06} and \citet{Sahai07}.

On the other hand, no CO emission is seen here at high velocity around the
SE-NW optical bipolar lobe. It has been argued that this bipolar lobe, like
the E-W bipolar lobe, is also a dense-walled structure with tenuous
interiors (or cavities) \citep{Sahai07} and could also be produced by a
bipolar post-AGB wind
\citep{Sanchez06,Sahai07}. 
Thus, this bipolar lobe could trace a bipolar outflow moving along the plane
of the sky, with the CO emission seen only at low velocity (within $\pm$ 10
\vkm{} from the systemic as seen in the channel maps) forming the
faint protrusions along the SE-NW bipolar lobe, as mentioned in the previous
section. The low inclination is also consistent with the fact that the SE
and NW components of the SE-NW bipolar lobe have almost the same brightness
in the optical.

\section{Modeling the torus-like envelope}\label{ssec:model}

As discussed, the two CO clumps seen at low velocity arise from a
torus-like envelope at a small inclination angle, with the nearside tilted
to the east.
In this section, we derive the physical properties, including the
kinematics, structure, density and temperature distributions, of this
torus-like envelope through modeling the two CO clumps with a
radiative transfer code. The envelope appears torus-like around the common
waist of the two
bipolar lobes, and it could arise from
a torus with or without the cavities (or holes)
cleared by the bipolar lobes, or a spherical shell with the cavities. For a
torus and a spherical shell, the number density of the molecular hydrogen in
spherical coordinates
$(r, \theta, \phi)$ can be assumed to be
\begin{equation}
n = n_0 \Bigl( \frac{r}{r_0}\Bigr)^{-2} \cos^p \theta,
\end{equation}
where $\theta$ is measured from the equatorial plane
of the envelope. 
Here $p=0$ for a spherical shell and $p > 0$ for a torus with a
half opening angle $\theta_0$ defined as
\begin{equation}
\cos^p \theta_0 = 0.5
\end{equation}
Also $r_0=$\arcs{1}, which is the
representative radius of the torus-like envelope. The envelope
has an inner radius $R_\textrm{in}$ and an outer radius
$R_\textrm{out}$. It is expanding radially at a velocity of $v_e$,
and thus the mass-loss rate (including helium) can be given by
\begin{equation}
\dot{M}=  1.4 m_{\scriptsize \textrm{H}_2}\int n v_e r^2 d\Omega
       =  2.8 \pi m_{\scriptsize \textrm{H}_2} n_0 v_e r_0^2 
          \int \cos^p \theta d\theta
\end{equation}
The cavities, if needed, are assumed to have a 
half-ellipsoidal (like paraboloidal) opening for
simplicity, with the semi-major axis and semi-minor axis determined from the bipolar lobes
in the optical.
Inside the cavities, the number density of the envelope is set to
zero.

In the envelope, the temperature profile is assumed to be
\begin{equation}
T = T_0 \Bigl(\frac{r}{r_0}\Bigr)^{-1}
\end{equation}
similar to that derived by \citet{Kwan82} for AGB envelopes, 
including gas-dust collisional heating, adiabatic cooling, 
and molecular cooling. 


In our models, radiative transfer is used to calculate the CO emission, with an
assumption of local thermal equilibrium. The relative abundance of CO to H$_2$ is assumed to be 4$\times10^{-4}$, as
in \citet{Sanchez06}. 
For simplicity, the line width is
assumed to be given by the thermal line width only. Also, the systemic
velocity $\vsys$ is assumed to be a free parameter.
The channel maps of the
emission derived from the models are used to obtain the integrated intensity
map, spectrum, and PV diagrams to be compared with the observation. Note
that the observed emission in the east, west, southeast, and northwest
protrusions at low velocity (Fig. \ref{12CO+HST}) is not from the envelope
and thus will not be modeled here. 
In our models, the free parameters are $\vsys$, $v_e$, $\Rin$, $\Rout$,
$p$ (torus or spherical shell), cavities (holes), inclination angle $i$, equator PA, $n_0$, and $T_0$.
Table \ref{ModParam} shows the best-fit
parameters for our models with different structures.
Our models all require $\vsys \sim$ 18.25$\pm1$ \vkm{} and
$v_e\sim$ 12.5$\pm1.5$ \vkm{} to fit the two velocity ends in the
PV diagrams and the spectrum. They require R$_{in}\sim$ \arcsa{0}{7} (3430
AU) and $R_\textrm{out} \sim$ \arcsa{1}{6} (7840 AU) to match the emission
peak positions. They also require $T_0$ to be $\sim$ 28 K to match the
emission intensity, resulting in an overall characteristic (or mean)
temperature of $\sim$ 23 K for the envelope. This characteristic temperature
is slightly higher than the values found in the more extended envelope by
\citet{Sanchez06} and \citet{Sahai07} at lower resolutions. This is
reasonable, because the envelope temperature is expected to be higher near
the source. In the following, we discuss our different models in detail.



\subsection{A Torus without cavities \label{Sec:pureT}}

First we check if a simple torus model can reproduce the two clumps (Figure
\ref{fig:NHTorusInc15}). In this model, the equator of the torus is set to
be at P.A.=15\degree, to be aligned with the two clumps. As expected, the
torus is required to have a small inclination angle of $\sim$ 15\degree{}
and a small half opening angle of
$\theta_0 \sim$ 24\degree{} with $p=8$. This model can reproduce
reasonably well the compactness of the two clumps. It can also reproduce the
ring-like PV structures, and the less and less emission around the systemic
velocity in the PV diagrams from cut A to cut D. Note that the PV diagram
cut along the equatorial plane shows a ring-like structure with two C-shaped
peaks at the two high velocity ends facing to each other, because the
emission is symmetric about the equatorial plane. For the PV diagrams with a
cut away from the equatorial plane, the blueshifted emission shifts to the
southeast and the redshifted emission shifts to the northwest, because the
torus is inclined with the nearside tilted to the southeast. For the
spectrum, the model can produce the double-peaked line profile, however, it
produces higher intensity than the observed around the systemic velocity
(Fig.
\ref{fig:NHTorusInc15}b).



\subsection{A Torus with a E-W pair of cavities \label{Sec:TorusEW}}

Here we add a E-W pair of cavities to the above simple torus model and see if
we can reproduce the spectrum better by reducing the intensity around the
systemic velocity (Figure \ref{fig:TorusEWp8}). The E-W pair of cavities are assumed to have a
half-ellipsoidal opening with a semi-major axis of \arcs{2} and a semi-minor
axis of \arcs{1}, as measured from the HST image (Fig.
\ref{12CO+HST}). 
The E-W pair of cavities have a P.A.=88\degree{} and can
have an inclination angle from 10\degree{} to 30\degree{}, with the
western cavity titled toward us, as suggested in \citet{Sanchez06}. 
These moderate inclinations are consistent with the
optical image that shows the eastern lobe is slightly fainter than the
western lobe, and they all give similar results in our model.
 The model results here are similar to those of the simple
torus model above, with some slight differences. This is because the E-W
pair of cavities are almost aligned with the poles of the torus, and
removing the tenuous material in the poles does not change much the model
results. The cavities remove slightly the blueshifted emission in the west
and redshifted emission in the east, mainly at low velocity, as seen in the
spectrum. In addition, unlike that in the torus model, the PV diagram cut
along the equatorial plane shows a ring-like structure with two
sickle-shaped peaks, instead of two C-shaped peaks, at the two high velocity
ends facing to each other (comparing Figure
\ref{fig:NHTorusInc15}c and Figure \ref{fig:TorusEWp8}c). This is because
that, along that cut direction, the E-W pair of cavities remove the
redshifted material in the north, and blueshifted emission in the south.


\subsection{A Torus with 2 pairs of cavities \label{Sec:T4h}}

Here we add one more pair of cavities, the SE-NW pair, into our model
(Figure \ref{fig:ModTorus}). Again,
this pair of cavities are assumed to have a half-ellipsoidal opening
but with a semi-major axis of \arcs{2} and a semi-minor axis of
\arcsa{0}{8}, as measured from the HST image (Fig. \ref{12CO+HST}). This
pair of cavities have a P.A.=$-40$\degree{} and are assumed to have an
inclination of 0\degree{} (i.e., in the plane of the sky). Adding this pair
of cavities will rotate the two clumps counterclockwise about the source.
Therefore, the torus in this model is required to have its equator at a
smaller P.A. of 10\degree{} in order for the two clumps to appear at
P.A.=15\degree{}. As a result, the underlying torus could be initially
more perpendicular to the E-W bipolar lobe. Also, a higher density is
required to reproduce the same amount of flux in the clumps (see Table
\ref{ModParam}).

By removing the material along the SE-NW bipolar lobe, this model can now
reproduce the required dip around the systemic velocity in the double-peaked
spectrum. This is because that the SE-NW pair of cavities, being in the
plane of the sky, remove the low-velocity emission preferentially. This
model can also reproduce the ring-like PV structures, and the less and less
emission around the systemic velocity in the PV diagrams from cut A to cut
D. Note that, in the PV diagram cut along the equatorial plane, the model
shows two sickle-shaped peaks, which could be somewhat different from the
observation in detail. This detailed difference could be due to localized
excitation effect near the base of the E-W cavities and should not affect our main conclusions on the envelope
properties.


\subsection{A Spherical shell with 2 pairs of cavities}\label{sec:sph}

A spherical shell with only the E-W pair of cavities can be ruled out
because it would produce two clumps that are exactly perpendicular to the
E-W bipolar lobe, inconsistent with the observation. Therefore, here we
check if a spherical shell with the same 2 pairs of cavities as above can
produce the two clumps in the observation.  We find that this model is quite
similar to the torus model with the 2 pairs of cavities, except that the two
clumps are slightly rotated and the emission is more extended perpendicular
to the torus-like structure, in between the cavities (comparing Figure
\ref{fig:ModSpherical}a to Figure \ref{fig:ModTorus}a).
This slight rotation is acceptable because
our model is simple and the two clumps are not resolved in our observation.
On the other hand, further observation that can separate the envelope from
the two bipolar lobes is needed to check if the emission
extending perpendicular to the torus-like structure in this model
is inconsistent with the observation.


\subsection{Model summary}

Our model results show that the two CO clumps could arise from either a torus
or a spherical shell, with
two pairs of cavities, one along the E-W bipolar lobe and one along the SE-NW
bipolar lobe. A pure torus model and a torus with only a E-W pair of cavities both
produce more than observed emission around the systemic velocity in the
spectrum. In addition, the SE-NW pair of cavities, being in the plane of the
sky, are essential to remove the low-velocity emission in our models.
All the models are optically thin. With the mean envelope temperature of 23
K and the model brightness temperature of $\sim$ 4 K, the mean optical depth is
$\sim$ 0.2. Therefore, in the model
spectra, the redshifted peak is only slightly brighter than the blueshifted
peak, as observed.
The mass-loss rate of the AGB wind derived from our two best models is $\sim
2-3\times10^{-5}$ \smyr{}. In the shell model, the mass-loss rate is about
three times smaller than that derived from a similar shell model by
\citet{Sanchez06}, which was found to be $\sim$ 10$^{-4}$ \smyr{}. This is
probably because of the following reasons: (1) our inner radius is smaller
than theirs, and the emission is more efficient in the inner region where
the density and temperature are both higher; (2) a large amount of missing
flux in our observation; and (3) their envelope emission seen at low
resolution could be significantly contaminated by the outflows. 
Therefore, the mass loss rate derived here could be a lower limit.
Alternatively, the mass-loss rate might have decreased with time.



\section{Discussion}\label{sec:Discuss}


\subsection{The expanding torus-like envelope}

\begin{figure}
\centering
\scalebox{0.8}{\includegraphics[angle=-90]{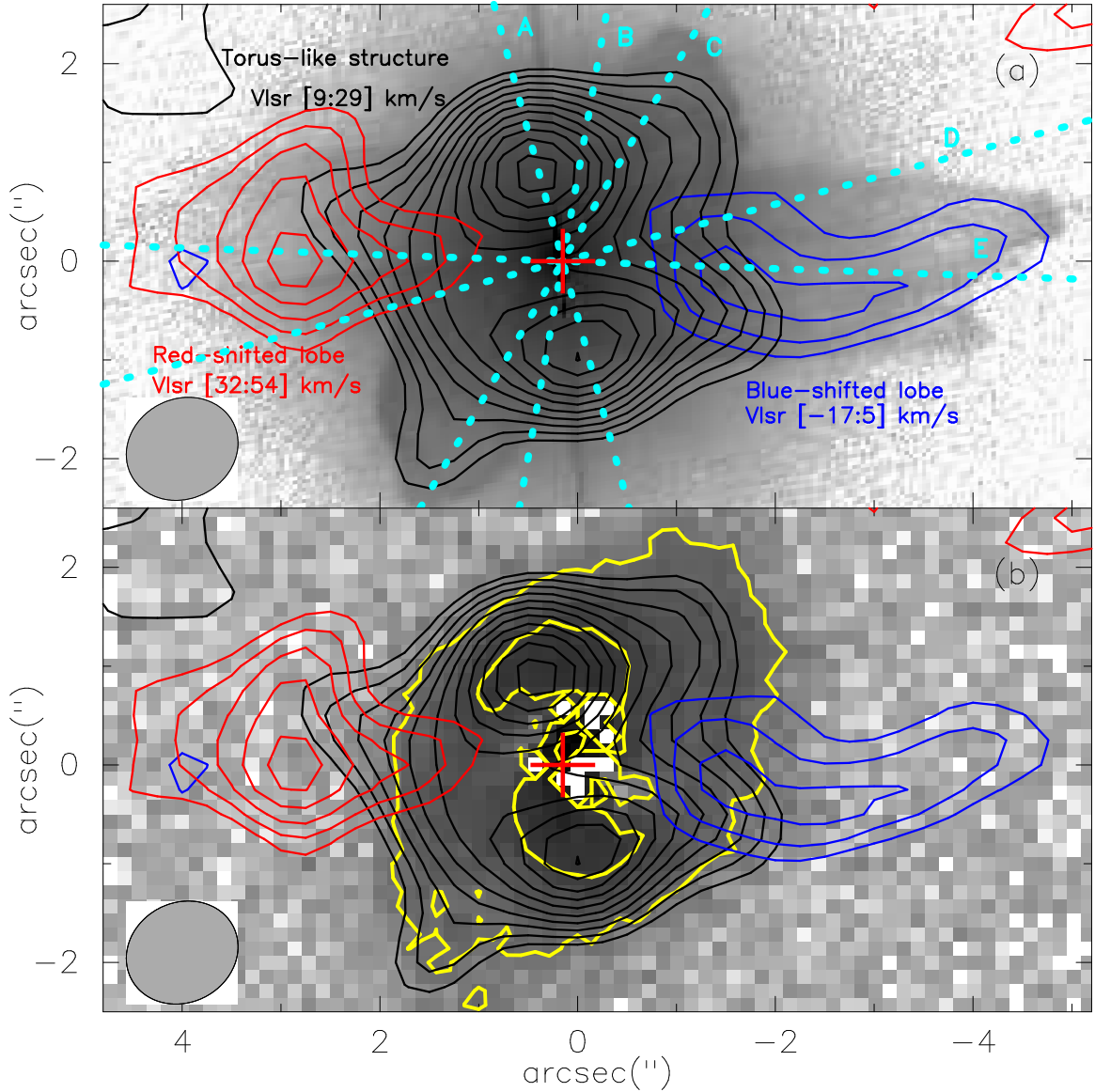}}
\figcaption[]
{CO J=2-1 contours of IRAS 19475+3119 superposed on top of (a) 
the HST image obtained with F606W filter \citep{Sahai07}
and (b) the NIR image \cite[with yellow contours,][]{Gled01}.
The central source position is marked by a ``+". The synthesized beam 
of the CO observation is shown in the left corner, with a size of 
\arcsa{1}{16}$\times$\arcsa{1}{01}.
The CO intensity maps integrated
over three selected LSR velocity intervals show the torus-like
structure (black contours), redshifted (red contours) 
and blueshifted (blue contours) outflow lobes.
Contour levels all start from 2$\sigma$ with a spacing of 1$\sigma$, where
$\sigma$ is 0.45 Jy beam$^{-1}$ \vkm{}.
Labels A--E show
the cuts for the position-velocity (PV) diagrams in Figure \ref{12CO-PVobs}.
\label{12CO+HST}
}
\end{figure}

\begin{figure}
\centering
\epsscale{0.9}
\scalebox{0.9}{\includegraphics{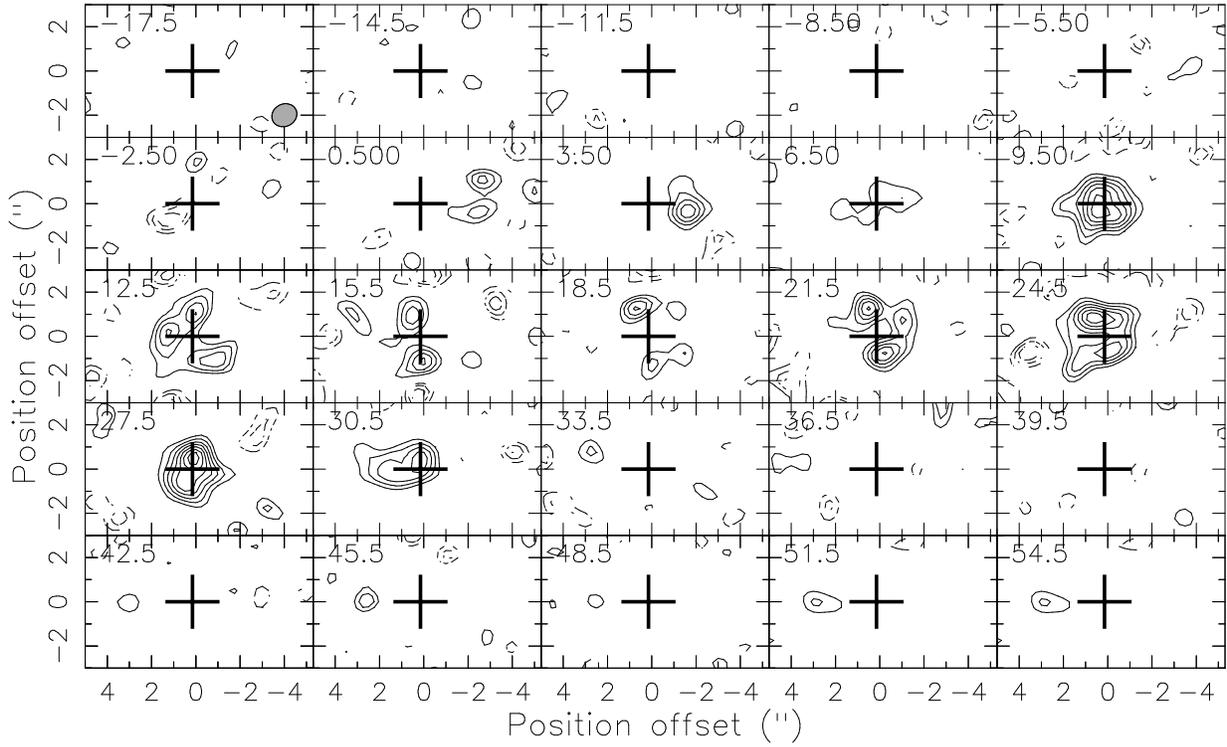}}
\figcaption[]
{Channel maps of CO J=2-1 emission in IRAS 19475+3119. 
The velocity width is 3.0 \vkm{}. The systemic velocity is 18.25$\pm$1 \vkm{},
as found in our model.
The contour levels start from $2\sigma$ with a spacing of
1$\sigma$, where $\sigma=$0.06 \Jyb{}.
The stellar position is marked by a cross.
The synthesized beam is shown in the first channel map at the lower right
corner. The velocity in each
channel is shown in the top left corner.
\label{Fig:Cha12CO}
}
\end{figure}

\begin{figure}
\centering
\epsscale{1.0}
\scalebox{0.65}{\includegraphics[angle=-90]{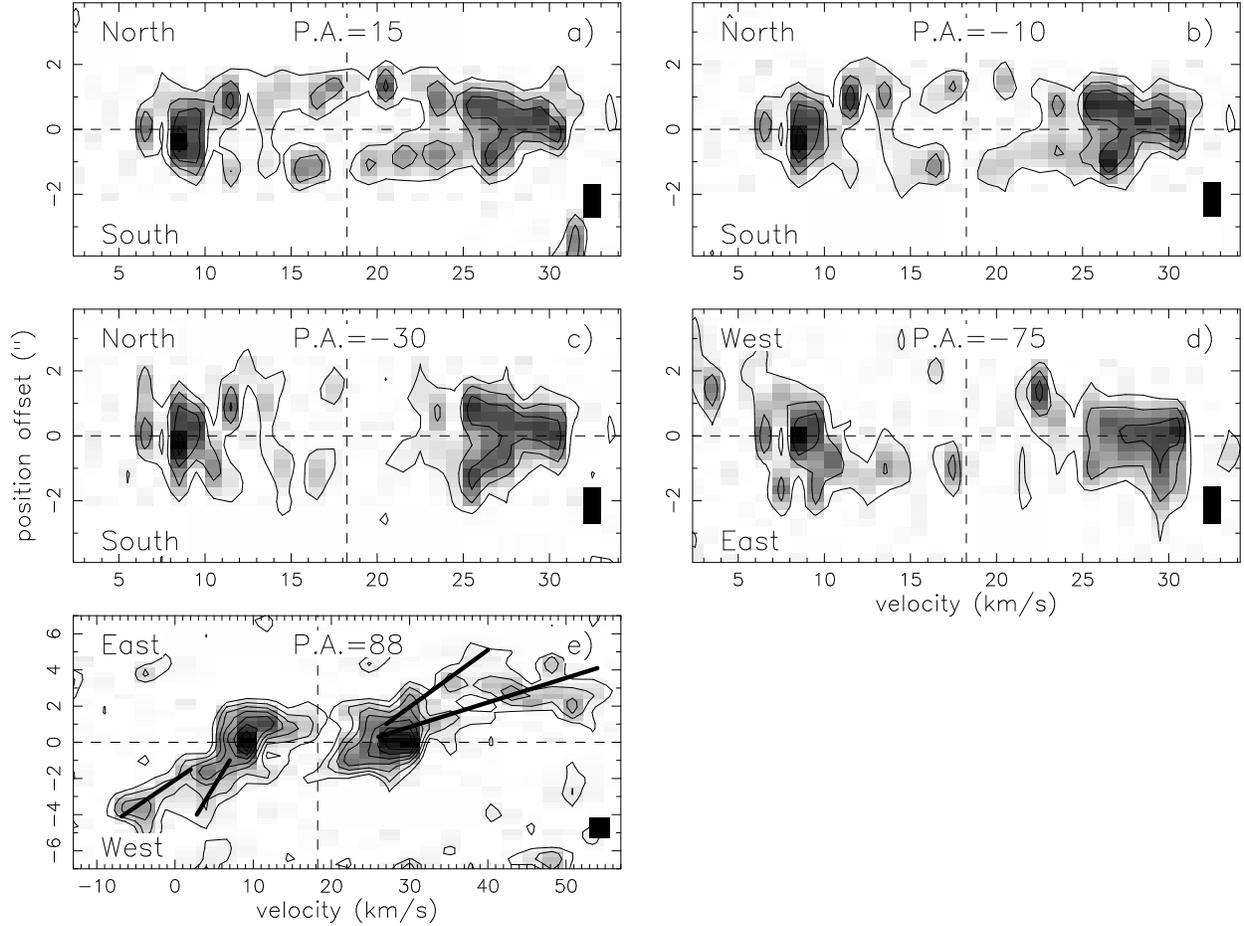}}
\figcaption[]
{PV diagrams of the CO emission, with the cut directions shown in Figure
\ref{12CO+HST}a. The contour levels start from 2$\sigma$ with a
spacing of 2$\sigma$, where $\sigma=0.06$ \Jyb{} in (a)-(d) and 0.03 \Jyb{}
in (e).
The spatial and velocity resolutions are shown as
a rectangle in the bottom right corner.
(a) Cut A (along P.A.=15\degree) goes through the emission peaks
of the two low-velocity clumps, showing a ring-like PV structure.
(b) Cut B (along P.A.=$-$10\degree) goes along the axis half way to the
edges of the clumps. It 
also shows a ring-like structure but with a weaker emission around the
systemic velocity.
(c) Cut C (along P.A.=$-$30\degree) goes through the edges of the clumps,
showing a breakup of the ring-like PV structure.
(d) Cut D (along P.A.=$-$75\degree) goes perpendicular to the axis
connecting the two clumps, showing the emission shifted to the east on the
blueshifted side and to the west on the redshifted side.
(e) Cut E (along P.A.=88\degree) goes along the high-velocity
component in the east-west direction,  showing a velocity increasing with the distance
from the source for the outflow, and a hint of bifurcation of the PV
structure (solid lines).
\label{12CO-PVobs}
}
\end{figure}

\begin{figure}
\centering
\scalebox{0.5}{\includegraphics[angle=-90]{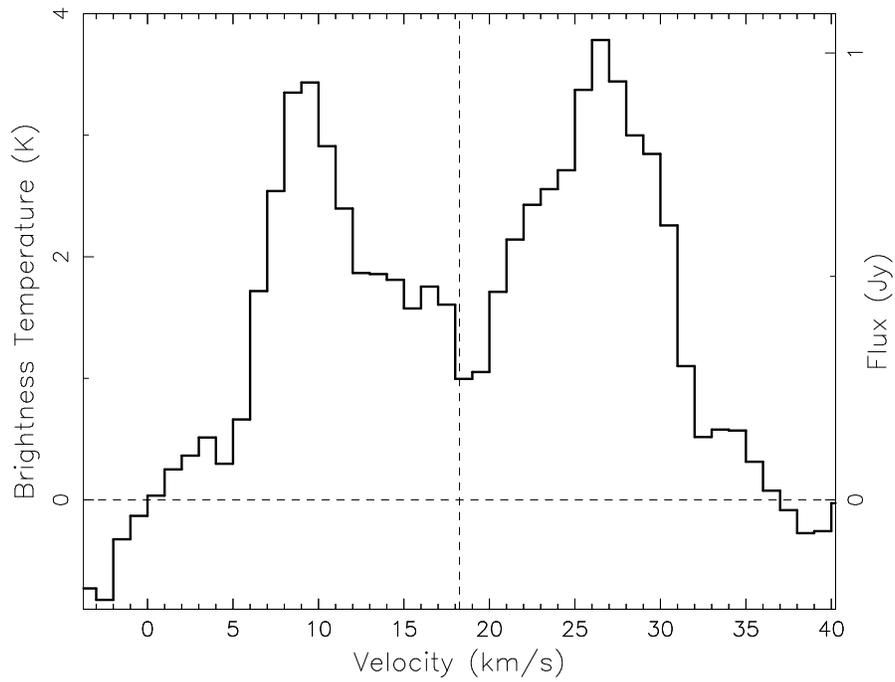}}
\figcaption[]
{CO spectrum toward the center position, averaged over a region with a
diameter of \arcsa{2}{5}. Note that a bigger region will include emission from
the two bipolar lobes. The vertical dashed line shows the systemic velocity. The
spectrum shows more emission on the redshifted side than the blueshifted
side, as expected for an expanding envelope. 
\label{spectrum}
}
\end{figure}

\begin{figure}
\centering
\epsscale{1.1}
\scalebox{0.7}{\includegraphics[angle=-90]{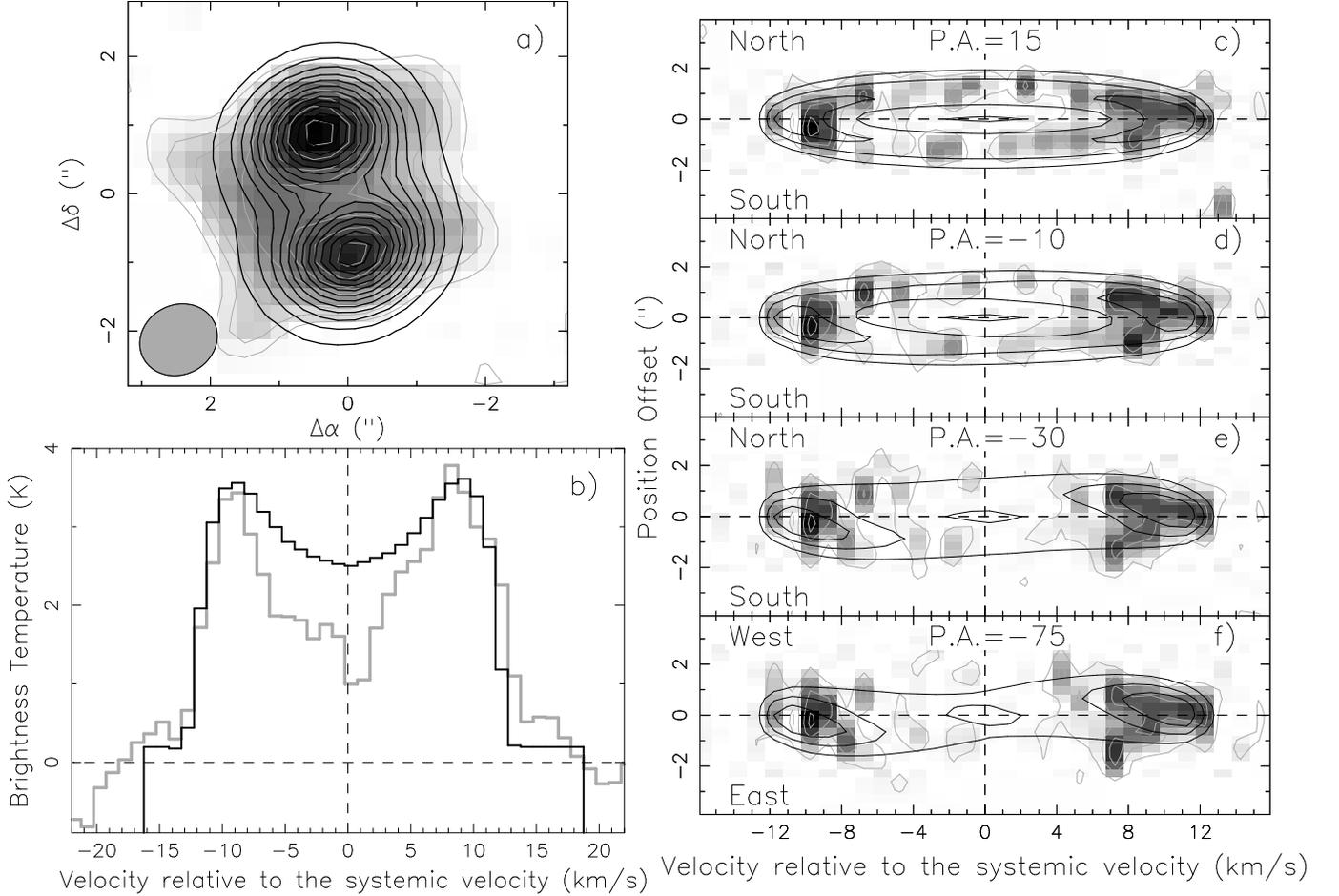}}
\figcaption[]
{Comparison of our torus model with the observed torus-like envelope in (a)
integrated intensity map as shown in Figure \ref{12CO+HST},
(b) spectrum as shown in Figure \ref{spectrum}, and (c)-(f) PV diagrams
as shown in Figure \ref{12CO-PVobs}a-d.
Black contours and spectrum
are from the model. The best fit model requires $p=8$ and 
an inclination of $\sim$15\degree.
\label{fig:NHTorusInc15}
}
\end{figure}

\begin{figure}
\centering
\epsscale{1.1}
\scalebox{0.7}{\includegraphics[angle=-90]{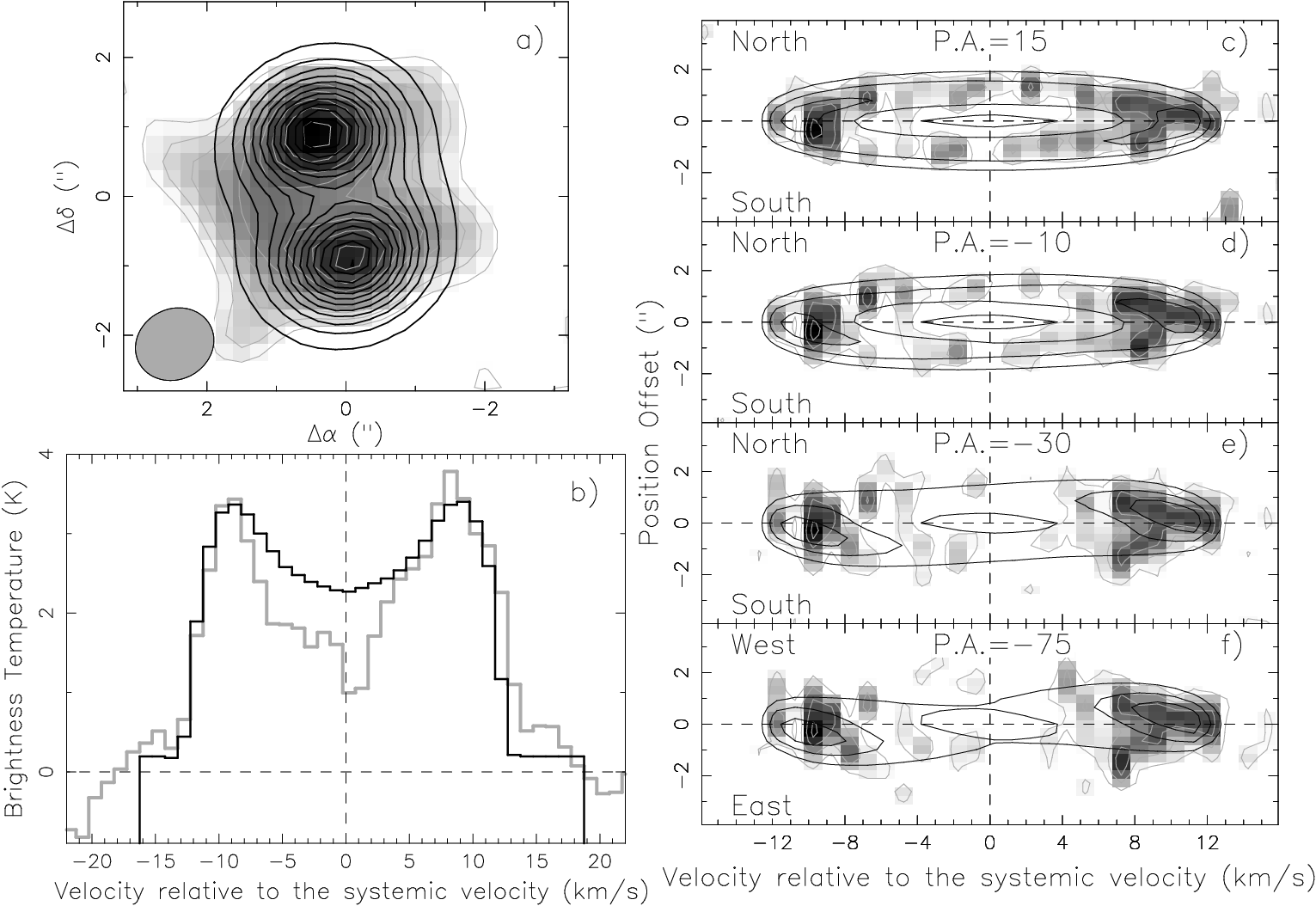}}
\figcaption[]
{Same as Figure \ref{fig:NHTorusInc15} but for the torus + EW pair of
cavities model.
\label{fig:TorusEWp8}
}
\end{figure}

\begin{figure}
\centering
\epsscale{1.1}
\scalebox{0.7}{\includegraphics[angle=-90]{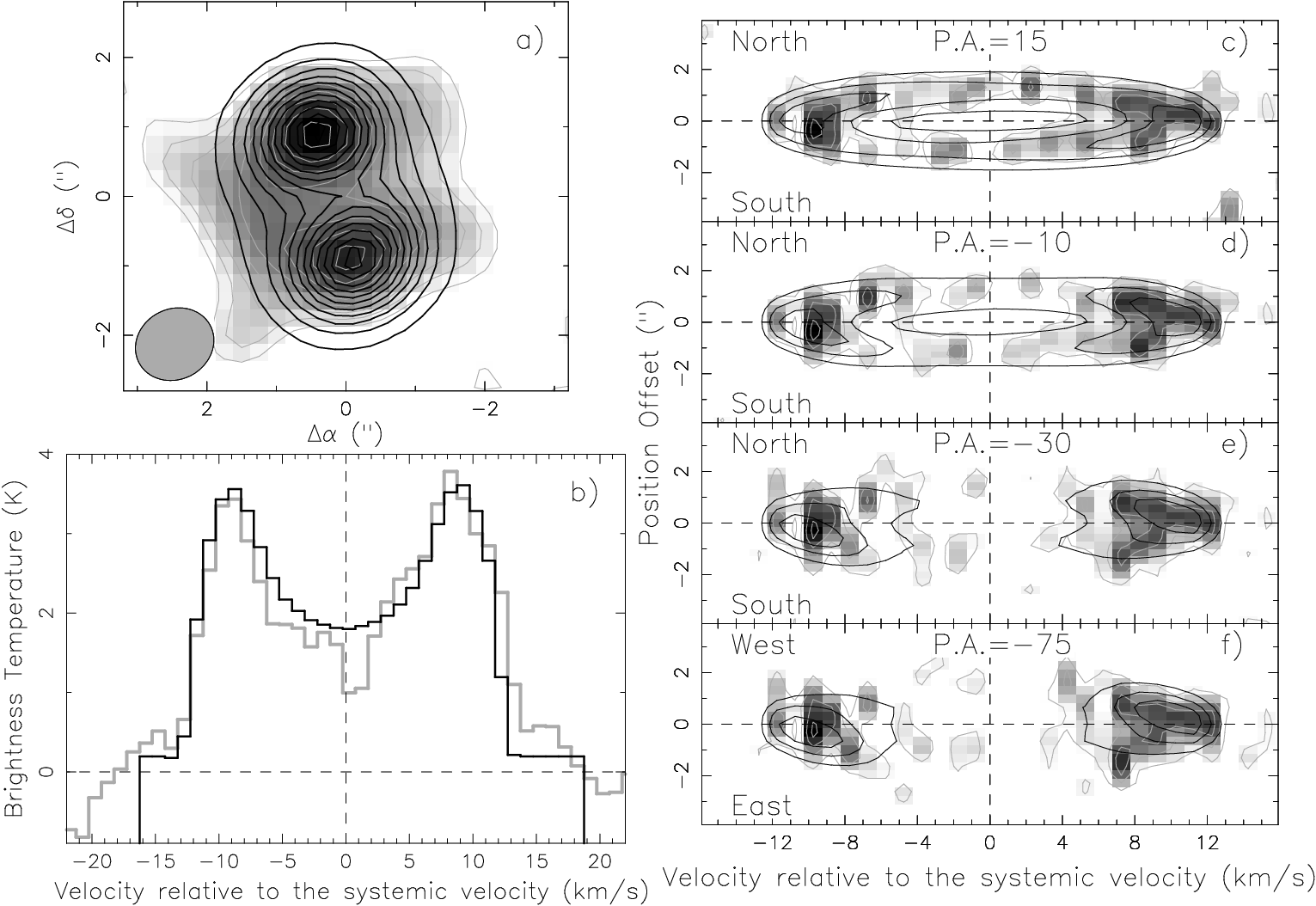}}
\figcaption[]
{Same as Figure \ref{fig:NHTorusInc15} but for the torus + 2 pairs of
cavities model.
\label{fig:ModTorus}
}
\end{figure}

\begin{figure}
\centering
\epsscale{1.1}
\scalebox{0.7}{\includegraphics[angle=-90]{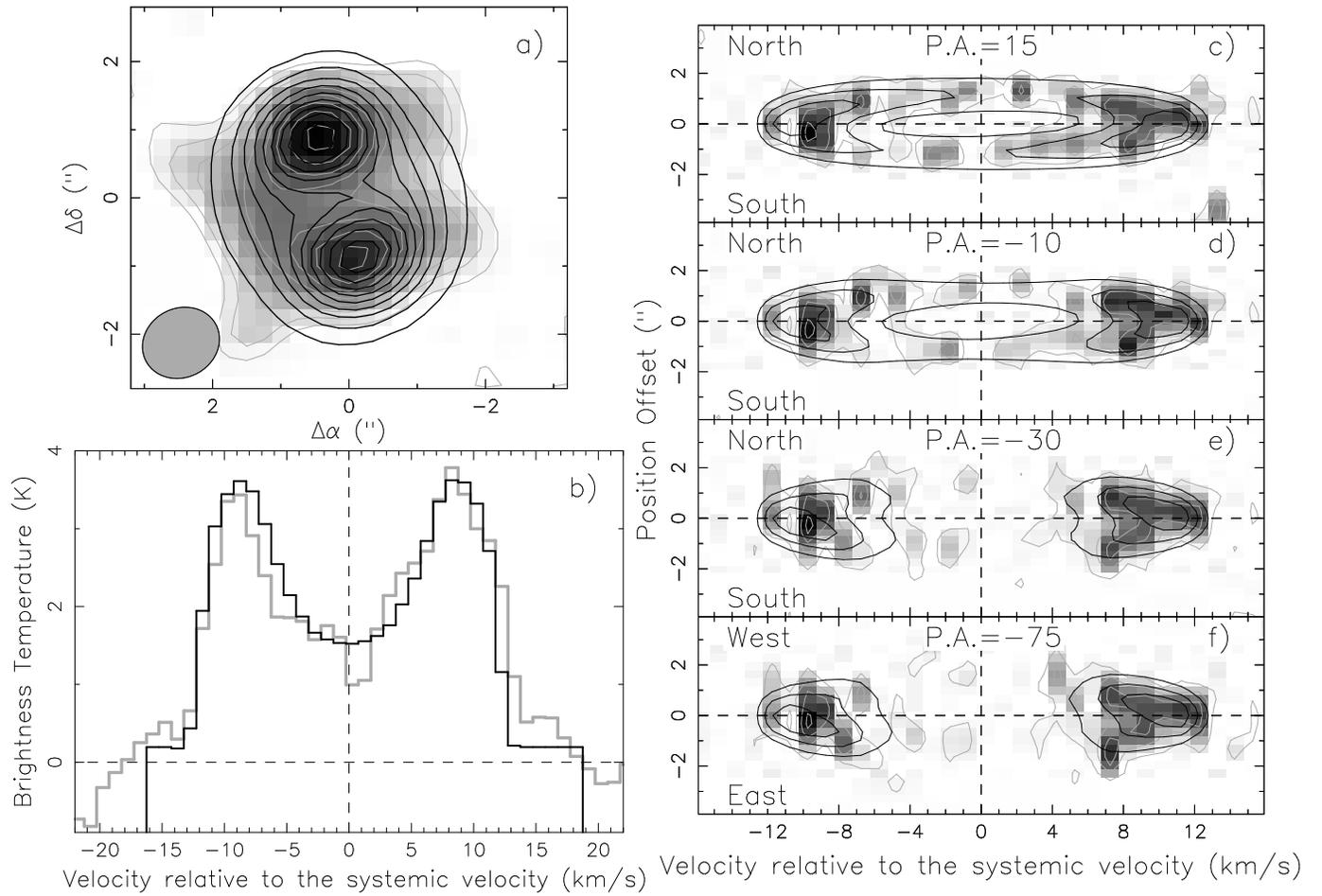}}
\figcaption[]
{Same as Figure \ref{fig:NHTorusInc15} but for the spherical shell + 2
pairs of cavities model.
\label{fig:ModSpherical}
}
\end{figure}

In our observation, two clumps are seen in CO at the common waist of the two
optical bipolar lobes, spatially coincident with the two peaks in the NIR
polarization image (Figure \ref{12CO+HST}), tracing the dense envelope
material. These two clumps are also seen at higher resolution by
\citet{Castro2010}, but not spatially resolved at lower resolution in
\citet{Sanchez06}. From our models, they are found to trace the two
limb-brightened edges of an expanding torus-like envelope around the source,
with an inclination angle of
$\sim$ 15\degree{}. This torus-like envelope corresponds to the ring-like
envelope found at higher resolution at a similar inclination angle in
\citet{Castro2010}. At a lower resolution of $\sim$ \arcs{2}, 
two different clumps were seen in CO
but aligned with the SE-NW optical bipolar lobe
\citep{Sanchez06} as well as the SE-NW NIR nebula. 
Those clumps are also seen here but appear
as the two faint protrusions
along the SE-NW bipolar lobe (Fig.\ref{12CO+HST}a). As mentioned before,
the two optical bipolar lobes, both E-W and SE-NW, have been found to be
dense-walled structures with tenuous interiors (i.e., cavities)
\citep{Sahai07} and these cavities are indeed required in our models. Thus,
our protrusions or their clumps along the SE-NW bipolar lobe
likely trace the unresolved dense wall materials
(i.e., swept-up material) of that bipolar lobe as suggested in
\citet{Sanchez06} and
\citet{Sahai07}.
They only appears at low velocity likely because the SE-NW bipolar lobe is
lying in the plane of the sky, as supported by the similar brightness of the
SE and NW components of the bipolar lobe, in both the HST optical and the
NIR images. They are faint here in our CO observation probably because their
CO emission merges with that of the extended circumstellar envelope and
halo, and is thus mostly resolved out. Two other
protrusions are also seen at low velocity along the E-W bipolar lobe (Fig.
\ref{12CO+HST}a), tracing the unresolved dense wall materials of the bipolar
lobe.

\subsection{The shaping mechanism and its consequences}

The PPN 19475+3119 is not spherically symmetric, with a torus-like envelope
at the common waist of the two bipolar lobes.  The torus-like envelope is
expanding at a speed of $\sim$ 12.5 \vkm{}, and thus has a dynamical age of
$\sim$ 1900 yr, with a radius of $\sim$ 5000 AU ($\sim$ \arcs{1}). The
E-W bipolar lobe has a mean inclination of $\sim$ 20\degree{}, so the
mean deprojected velocity is $\sim$ $27/\sin(20) \sim$ 80 \vkm{} at the tips and the mean
deprojected distance of the tips is
$\sim$ 21000 AU ($\sim$ \arcsa{4}{3}). Thus,
the dynamical age of the E-W bipolar lobe is estimated to be
$\sim$ 1200 years ($\pm$400 years), more than half of that of the torus-like
envelope. The SE-NW bipolar lobe is about half the length of the E-W bipolar
lobe and may have a similar
dynamical age to the E-W bipolar lobe within a factor of two. 
The bipolar lobes and torus-like envelope are
believed to be physically related.  
The envelope
is believed to be the AGB envelope formed by the late AGB wind. 
The two bipolar lobes could trace two
bipolar outflows produced by two distinct bipolar post-AGB winds ejected in
two different directions, as proposed in \citet{Sanchez06} and
\citet{Sahai07}.
The densed-wall structures with tenuous interiors (cavities) of the bipolar
lobes could result from the interactions between the post-AGB winds and the
envelope as shown in hydrodynamical simulations \citep{Lee03,CRL618b}. Also,
according to our models, the envelope, originally either a torus or a
spherical shell, is disrupted by the two bipolar lobes or outflows, and it
appears as it is now because of the disruptive interactions.




Could the torus-like envelope originally be a torus, like the dense
equatorial waist structure often seen in other PPNe 
\citep{Volk07,Sahai08}?
If so, it could be due to an equatorially enhanced mass loss in the late AGB
phase \cite[see the review by][]{Balick02}. Such a dense waist, where large
grains can grow, seems to be needed to account for the
millimeter/submillimeter excess toward this source, as argued by
\citet{Sahai07}. This toroidal envelope may help collimate the E-W
bipolar outflow in its pole direction. However, it is hard for the toroidal
envelope to collimate two bipolar outflows that are almost perpendicular to each
other, such as those in this PPN. Instead, the envelope appears
to be disrupted by the two outflows, as discussed above.


On the other hand, could the torus-like envelope be a remnant of a
spherical AGB shell in the inner part? The torus-like envelope has an
expansion velocity similar to that of the surrounding halo
\citep{Sahai07}, 
and thus may have the same origin as the halo,
which is the spherical circumstellar shell formed by an isotropic AGB wind in the
past. As discussed before, a
spherical shell model with 2 pairs of cavities along the two bipolar lobes
can produce a torus-like envelope as observed. Thus, the torus-like envelope
could indeed be the remnant of the AGB shell in the inner part. Note that
\citet{Sanchez06} have also tried a similar spherical shell model but only
with a pair of cavities along the E-W bipolar lobe. As mentioned in Section
\ref{sec:sph}, however, this model would produce two clumps that are
perpendicular to the E-W bipolar lobe, but not at the common waist of the
two bipolar lobes as observed.
A similar model with 2 pairs of cavities has been adopted to explain the clumpy hollow
shell morphology of the Egg Nebula by \citet{DinhV09}. In that PPN, the 
spherical AGB envelope is also disrupted by the outflows. In this PPN,
the spherical shell is disrupted more dramatically, leaving behind a
torus-like envelope. 



The envelope, originally either a torus or a spherical shell, will not be
able to collimate the two bipolar outflows, as discussed. Therefore, the two
collimated outflows in this PPN must be produced by two post-AGB
winds that are intrinsically collimated or jets, as proposed in other 
PPNe and young PNe
\citep{ST98,Sahai01,Lee03,Sanchez06}.
One popular model to produce a collimated jet requires a binary companion to
accrete the material from the AGB star and then launch the jet \cite[see the
review by][]{Balick02}.
\citet{Gled01} also suggested that a binary interaction is needed to produce
the two-armed spiral structure seen in the NIR polarization image of this
PPN. The binary interaction can also transform the isotropic mass
loss of the AGB wind into an equatorial enhanced mass loss 
\citep{Mastrodemos98}, producing a
toroidal envelope as needed in our model. However, 
it is unclear how two jets can be launched in this PPN
to produce the two bipolar outflows \cite[see][for various
possibilities]{Sahai07}.

\section{Conclusions}
IRAS 19475+3119 is a quadrupolar PPN, with two bipolar lobes, one in the
east-west (E-W) direction and one in the southeast-northwest (SE-NW)
direction. The E-W bipolar lobe is known to trace a bipolar outflow and it
is detected at high velocity. The SE-NW bipolar lobe appears at low
velocity, and could trace a bipolar outflow moving in the plane of the sky.
Two compact clumps are seen at low velocity around the common waist of the
two bipolar lobes, spatially coincident with the two emission peaks in the
NIR, tracing dense envelope material. They are found to trace the two
limb-brightened edges of a torus-like circumstellar envelope, expanding away
at $\sim$ 12.5 \vkm{}.
This torus-like envelope can be reproduced reasonably well with either a
torus or a spherical shell, both with two pairs of cavities along
the two bipolar lobes. Here, the torus could come from an equatorial
enhanced mass loss and the spherical shell could come from an isotropic mass
loss, both in the late AGB phase.  In either case, the circumstellar envelope
appears to be disrupted by the two bipolar outflows. 

\acknowledgements
We thank the referee for the valuable comments.
M.-C. Hsu appreciates valuable discussion with Chun-Hui Yang, especially in
the development of the radiative transfer code. C.-F. Lee and M.-C. Hsu
are financially supported by the NSC grant NSC96-2112-M-001-014-MY3.


\begin{deluxetable}{lcccc}
\tabletypesize{\normalsize}
\tablecaption{The best fit parameters for our models \label{ModParam}}
\tablewidth{0pt}
\tablehead{\colhead{Parameters} & 
\colhead{Torus} & \colhead{Torus}  &
\colhead{Torus} & \colhead{Shell Model}\\
\colhead{} & 
\colhead{no cavities} & \colhead{+ EW cavities}  &
\colhead{+ 2 pairs of cavities} & \colhead{+ 2 pairs of cavities}}
\startdata
$p$             & 8   & 8  & 8  & 0  \\
Inclination $i$ & 15  & 15 & 15 & -- \\
equator P.A.    & 15  & 15 & 10  & -- \\
$n_0$ (cm$^{-3}$)  & $7.5\times 10^{3}$  &  $7.5\times 10^{3}$  &  
                     $9.0\times 10^{3}$  &  $4.3\times 10^{3}$  \\
$T_0$ (K)      &  28 & 28 & 28 & 28 \\
$v_e$ (\vkm{}) & 12.5 & 12.5 & 12.5 & 12.5  \\
$\dot{M}$ (\smyr{}) &  $1.9\times10^{-5}$ &   $1.9\times10^{-5}$  &
                       $2.3\times10^{-5}$ &   $2.7\times10^{-5}$  \\
\enddata
\tablecomments{The torus model with p=8 has a half opening angle 
of $\theta_0 \sim 24$\degree. Here the 2 pairs of cavities include the E-W
pair and SE-NW pair of cavities. The E-W pair of cavities have a
P.A.=88\degree{} and a mean inclination angle of $\sim$ 20\degree{}, with the
western cavity titled toward us. The SE-NW pair of cavities have a
P.A.=$-40$\degree{} and are assumed to have an inclination of 0\degree{}.}
\end{deluxetable}

\clearpage

\end{document}